\documentclass[conference]{IEEEtran}
\IEEEoverridecommandlockouts
\usepackage{cite}
\usepackage{amsmath,amssymb,amsfonts}
\usepackage{algorithmic}
\usepackage{graphicx}
\usepackage{textcomp}
\usepackage{xcolor}
\usepackage{booktabs}
\usepackage{multirow}
\usepackage{subcaption}
\def\BibTeX{{\rm B\kern-.05em{\sc i\kern-.025em b}\kern-.08em
    T\kern-.1667em\lower.7ex\hbox{E}\kern-.125emX}}

\begin{document}

\title{IFS: Information Flow Structure for Multi-agent Ad Hoc System}

\author{
\IEEEauthorblockN{Yanqing Fu\textsuperscript{*}}
\IEEEauthorblockA{\textit{College of Electronics and Information Engineering} \\
\textit{Tongji University}\\
Shanghai, China \\
fuyanqing159@163.com}

\and
\IEEEauthorblockN{Chenrun Wang\textsuperscript{*}}
\IEEEauthorblockA{\textit{College of Electronics and Information Engineering} \\
\textit{Tongji University}\\
Shanghai, China \\
1015239683@qq.com}

\and
\IEEEauthorblockN{Chao Huang}
\IEEEauthorblockA{\textit{College of Electronics and Information Engineering} \\
\textit{Tongji University}\\
Shanghai, China \\
csehuangchao@tongji.edu.cn}

\and
\IEEEauthorblockN{Zhuping Wang}
\IEEEauthorblockA{\textit{College of Electronics and Information Engineering} \\
\textit{Tongji University}\\
Shanghai, China \\
elewzp@tongji.edu.cn}
}

\renewcommand{\thefootnote}{\fnsymbol{footnote}}

\maketitle

\footnotetext[1]{These authors contributed equally to this work.}

\renewcommand{\thefootnote}{\arabic{footnote}}

\begin{abstract}
Multi-agent ad hoc systems are dynamic collaborative systems in which multiple autonomous agents must cooperate with both known and unknown teammates in open environments, without relying on pre-coordinated strategies. These systems operate under conditions of uncertainty and partial observability, where team composition, agent behaviors, and environmental factors may change during execution. Through an analysis of information flow in such systems, we identify two key limitations in existing research: insufficient information flow and limited information processing capacity. To address these issues, we propose an information flow structure for multi-agent ad hoc systems (IFS), which tackles these challenges from the perspectives of communication and information fusion. Experimental results in StarCraft II demonstrate that IFS significantly improves both information flow and processing capacity, while exhibiting strong generalization capabilities and outperforming baseline methods in complex ad hoc teamwork scenarios.
\end{abstract}


\section{Introduction} \label{section:introduction}

Multi-agent reinforcement learning (MARL) is a branch of machine learning in which multiple autonomous agents learn to make sequential decisions through interactions within a shared environment \cite{2018ai, 2023tizero, 2018qmix} . A prominent area within MARL is cooperative MARL (CMARL), where agents must learn to coordinate their behaviors to accomplish a common goal \cite{1994markov, 2005cooperative}. In such settings, all agents receive the same global reward signal, which depends on their joint actions. The central challenge is for the agents to learn policies that maximize this collective return, often necessitating the development of implicit or explicit communication protocols and the ability to model or predict other agents' behaviors for effective collaboration \cite{2019emergent, 2019graph, 2019qtran}.

A prominent recent research direction with strong practical implications is ad hoc teamwork (AHT), in which an autonomous agent must collaborate fluently with other agents to accomplish a common goal, without prior coordination or explicit strategy alignment \cite{2022aht, 2010aht}. This line of work has been extended to n-agent ad hoc teamwork (NAHT), generalizing the problem from a single agent cooperating with others to a team of multiple agents working together with external partners \cite{2024adhoc}. NAHT is particularly relevant in scenarios such as mixed human-robot teams, autonomous vehicles navigating among human-driven cars, and disaster response missions where predefined coordination protocols are unavailable. The central challenge lies in enabling agents to quickly adapt to unfamiliar teammates, infer their behaviors and intentions in real time, and effectively align their actions with collective objectives. This demands capabilities such as online reasoning, robust policy adaptation, and teammate modeling under uncertainty \cite{2022decision, 2021coach}. 

A multi-agent ad hoc system, as explored under the framework of NAHT, operates in open-system environments characterized by dynamic and unpredictable conditions \cite{2021open, 2023general}. In such settings, system components, participating agents, and interaction rules are not fixed and may change during operation. Unlike closed systems with clearly defined boundaries, open systems must persistently interact with and adapt to external elements. These include the introduction of new agents with unknown behaviors, dynamically shifting goals, and continuously evolving environmental dynamics \cite{2021formation, 2019action, 2021unmas}. This context introduces fundamental challenges such as non-stationarity, where the environment evolves from each agent's perspective, and partial observability, which confines agents to limited local views of the global state. Consequently, the central design objective is to develop agents that demonstrate robustness, generalize effectively to novel situations, and maintain adaptive performance when encountering unexpected conditions.

In this paper, we examine multi-agent ad hoc systems from an information flow perspective, focusing on how information moves among agents within the system and between internal and external agents. Since information serves as a prerequisite for system decision-making, how it flows and is processed within the system becomes critically important.

\subsection{Contribution}

We have conducted a thorough analysis of existing research on multi-agent ad hoc systems and identified two key limitations: insufficient flow of information within agents and the presence of system-imposed upper limits on information processing. To partially address these issues, we propose the information flow structure for multi-agent ad hoc system (IFS). The main contributions of this work are as follows:
\begin{enumerate}
\item We propose a communication protocol for controlled agents (CPCA) specifically designed for NAHT settings. This protocol aims to enhance information sharing among agents within the system while improving coordination capabilities with external agents. Building upon this protocol, we have designed a corresponding communication module (CM) within the IFS architecture.
\item To eliminate the need for preset limits on agent capacity during deployment, we introduce the concept of information fusion, which enables the transformation of variable-length data into a unified representation, thereby enhancing adaptability to open-system environments. Building on this concept, we have designed an information fusion module (IFM) and an information separation module (ISM) within the IFS.
\end{enumerate}

We evaluate the effectiveness of the proposed IFS through experimental testing in StarCraft II. Our results demonstrate that IFS significantly enhances information flow and processing while exhibiting strong generalization capabilities.

\subsection{Related work}

Centralized Training with Decentralized Execution (CTDE) has emerged as a prominent framework to address the non-stationarity challenge in multi-agent learning \cite{2018modelfree, 2017value}. In this paradigm, agents utilize centralized information during training to develop coordinated strategies, while executing decisions independently based solely on their local observations during deployment. This architecture effectively reconciles the need for strategic coordination during learning with practical demands for scalability and adaptability in real-world applications. As a result, CTDE has been widely incorporated into numerous multi-agent algorithms \cite{2021roma, 2020multi, 2021qplex, 2021qvalue}.

The CTDE framework is commonly employed in both CMARL and AHT. Existing AHT methods predominantly focus on scenarios involving a single controlled agent coordinating with other teammates. Typically, these approaches encode team behavior through neural network-based feature extraction, subsequently utilizing the resulting feature vector to determine the controlled agent's policy \cite{2021model, 2021bayesian}. Alternative methods enhance the controlled agent's adaptability by repeatedly training it with diverse teammate policies \cite{2021trajectory}. On the other hand, CMARL represents the opposite extreme, assuming full control over all agents without any need for ad hoc collaboration with unknown teammates. Common techniques in CMARL include credit assignment or role allocation to facilitate division of labor and cooperation among agents \cite{2018coma, 2021scaling, 2022ldsa}. Nevertheless, these methods fundamentally rely on predefined knowledge of all teammates, rendering them unsuitable for environments requiring spontaneous coordination with unfamiliar agents \cite{2020options, 2020other}.

Based on this, \cite{2024adhoc} pioneered the concept of NAHT by categorizing agents into controlled and uncontrolled types. Controlled agents are mutually known and governed by the algorithm, while uncontrolled agents remain unknown to the controlled agents, meaning that their goals, decision-making logic, and communication capabilities cannot be predetermined. Controlled agents must infer the intentions of uncontrolled agents solely through observations to coordinate and accomplish tasks. Accordingly, \cite{2024adhoc} introduced the policy optimization with agent modeling (POAM) algorithm, tailored to the characteristics of NAHT. However, as noted in \cite{2024adhoc}, this method was only experimentally applied to the problem and did not account for information flow among agents. In other words, controlled agents in POAM make decisions based solely on local observations. Since inferring other agents' intentions solely through external behavioral observations is inherently limited, the resulting collaboration among agents remains superficial.

Furthermore, when considering all controlled agents as a unified entity, they form a multi-agent ad hoc system. Such systems are typically open in practical applications, meaning that the number of agents in an ad hoc team is inherently variable. However, neural networks generally require fixed-length inputs, leading to the common practice of setting an upper limit on the number of agents and padding variable-length data to a uniform size. This approach is also adopted by \cite{2024adhoc} for feature extraction of agents in ad hoc teams. Clearly, this method constrains the system's openness and introduces two significant issues. First, when the number of agents in an ad hoc team exceeds the preset limit, the system must selectively discard some data. This raises the question of how to selectively remove data while preserving as much information as possible. Second, the order of agent data directly influences the representation after padding, creating the challenge of how to process differently ordered data while ensuring consistent outcomes.

\section{Background} \label{section:background}

\subsection{Multi-agent system}

A decentralized partially observable Markov decision process (Dec-POMDP) \cite{2002Markov} is a formal framework for modeling multi-agent systems where a team of agents must cooperate to achieve a common goal under two key constraints: decentralization and partial observability. It can be formally defined by the tuple 
$$
G = (I, S, A, T, O, r, \gamma).
$$
$I = \{1,\ldots, k\}$ is the set of $k$ agents. $S$ is the state space. $A = \prod_{i\in I} A_i$ is the set of joint actions, with $A_i$ being the action space of agent $i$. $T: S \times A \to \Delta(S)$ is the transition function\footnote{$\Delta(S)$ denotes the space of probability distributions on $S$.}, which maps a state and a joint action to a probability distribution over next states. $O = \prod_{i\in I} O_i$ is the set of joint observations, where $O_i$ is the private observation space of agent $i$. $r: S \times A \to \mathbb{R}$ is the common reward function shared by all agents. $\gamma \in [0,1)$ is the discount factor.

In MARL \cite{2023reinforcement-learning}, there are two additional important notations. $H = \prod_{i \in I} H_i$ is the set of joint histories, where $H_i$ is the history of agent $i$, recording its sequence of local observations and actions. $\pi = \prod_{i \in I} \pi_i$ is a joint policy, where $\pi_i: H_i \to \Delta(A_i)$ is the policy of agent $i$, mapping its individual history to a probability distribution over its actions.

In the MARL process, at timestep $t$, agent $i$ records its previous action $a_i^{t-1}$ and current local observation $o_i^t$ into its history $H_i^t$. It then uses its policy $\pi_i$ to generate a probability distribution over its action space $A_i$, and samples an action from this distribution as its current action $a_i^t \sim \pi_i(H_i^t)$. For the overall game $G$, the current actions of all agents are aggregated to form the joint action $a^t$. Based on the current state $s^t$, the system receives a reward $r_t = r(s^t, a^t)$. The state transition distribution $T(s^t,a^t)$ is then obtained from the transition function, and a new state is sampled as the next state $s^{t+1} \sim T(s^t,a^t)$.

The objective of reinforcement learning is to maximize the expected discounted cumulative return starting from any state $\mathbb{E}[\sum_{t = 0}^{\infty} \gamma^t r_t]$.

\subsection{N-agent ad hoc teamwork}

N-agent ad hoc teamwork (NAHT) requires an agent to cooperate effectively in teams that may include both known and unknown partners \cite{2024adhoc}. The "ad hoc" aspect signifies that these teams are formed on a temporary basis. The model can be represented as $(C,U)$, where $C$ denotes the set of controlled agents and $U$ denotes the set of uncontrolled agents. In other words, the agents in $C$ are mutually known to each other, while the agents in $U$ are unknown to the agents in $C$. In each task, $n \geq 1$ agents are randomly selected from $C$ to form $C'=\{c_1,\ldots,c_n\}$, and $m \geq 0$ agents are randomly selected from $U$ to form $U'=\{u_1,\ldots,u_m\}$. These selected agents then constitute the agent set $I=C' \cup U'$ of the multi-agent system $G$.

We need to train the policies of the agents in $C$. Denote the policy of agent $i \in C$ as $\pi_i(\cdot \mid \theta_i)$, where $\theta_i$ represents the parameters of the policy, which are the parameters to be trained. The objective of NAHT is to maximize the expected discounted cumulative reward of the multi-agent system under the above model\footnote{$\vert C \vert$ denotes the cardinality of the set $C$.} :
$$
\max_{\theta_1, \ldots, \theta_{\vert C \vert}} \mathbb{E}_{C',U'} \left[ \sum_{t = 0}^{\infty} \gamma^t r_t \right].
$$

In short, the core of NAHT lies in learning a set of policies for controlled agents, enabling them to collaborate effectively with other agents in ad hoc teams to accomplish tasks collectively. Notably, since these team formations are unknown in advance and the strategies of other agents are also unknown, NAHT must address the challenges inherent in open-system environments.

\section{Information flow structure} \label{section:methodology}

\begin{figure}[htbp]
\centering
\includegraphics[width=0.45\textwidth]{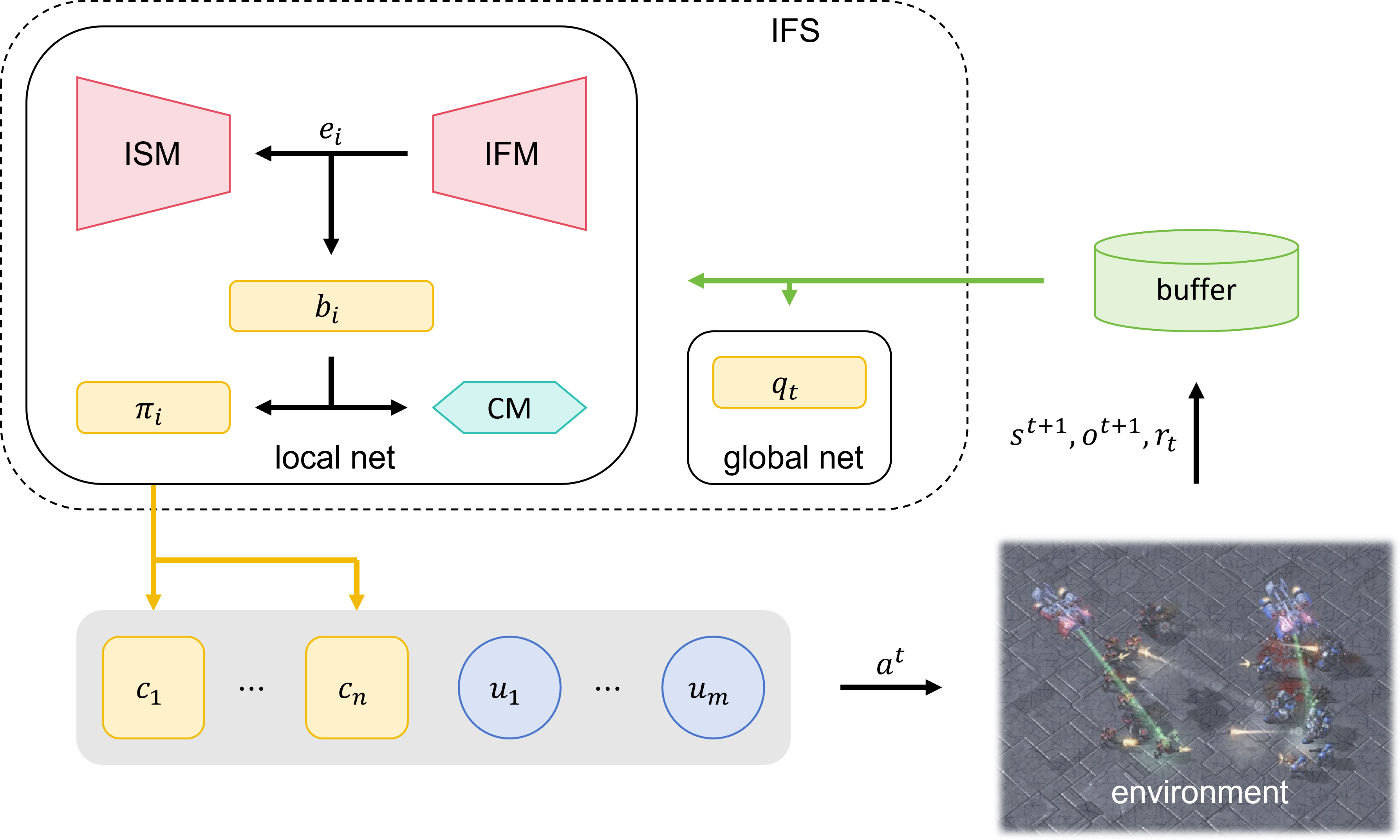}
\caption{Training process of IFS. The actor network receives local observation data and local communication data. It fuses this information into a representation $e_i$ through an information fusion module (IFM), then updates the controlled agent's internal state $b_i$. This internal state $b_i$ is used by a communication module (CM) to generate the agent's communication data, which is broadcast to nearby controlled agents. Simultaneously, $b_i$ is used to generate the agent's current policy $\pi_i$. During the training phase, the representation $e_i$ is also processed by an information separation module (ISM) to separate specific information for use as a reconstruction error. Additionally, the critic network receives global information to produce an action-value estimate $q_t$ for the current global state.}
\label{figure:overall}
\end{figure}

This section describes our proposed information flow structure for multi-agent ad hoc system (IFS). The core idea of this structure is to enhance the collaborative capability of controlled agents when forming ad hoc teams with either known or unknown agents by strengthening information flow. IFS relies on communication and information fusion to achieve more accurate inference of other agents' intentions and better adaptation to open-system environment. To reduce training overhead, we employ a parameter sharing mechanism, i.e., $\theta_1 = \ldots = \theta_{\vert C \vert}$. Figure~\ref{figure:overall} illustrates the training process of IFS.

\subsection{Overall system}

Our proposed IFS is a value-based reinforcement learning framework. Each controlled agent uses its local network to compute action-values for each action in its action set based on local observations, then selects an action as its policy according to a specific strategy. The global network evaluates the action-value based on the current joint action composed of all agents' actions. Therefore, the overall system's loss function is value-based and can be expressed as
\begin{equation}
L_{sys} = \sum_{t \geq 0} \vert r_t + \gamma \cdot \max_{a^{t+1}} q'(s^{t+1}, a^{t+1}) - q_t \vert^2,
\end{equation}
where $q'$ represents a copy of the global network $q$ and does not participate in parameter optimization during the training process. 

The system's loss function corresponds to the Double DQN (DDQN) loss function, which aims to improve the estimation of action-values for joint actions. By leveraging accurate action-values, each controlled agent can select appropriate actions as its policy based on these values, thereby enhancing the overall system's cumulative reward.

\subsection{Communication}

Information flow plays a crucial role in the system, as it determines how each component should operate. Each agent can acquire partial information through local observations. Typically, they can obtain information about other agents within their visible range, such as relative positions, agent types, and current states. While these observations allow agents to roughly assess the situations of others and make decisions accordingly, the information is generally limited to external characteristics of other agents. Such superficial data tends to be too generalized to accurately infer the intentions of other agents. As a result, collaboration among agents remains shallow, hindering the potential for deeper cooperation.

To partially address this issue, we propose a communication protocol for controlled agents (CPCA), which consists of two components. First, a local communication mechanism should be established among controlled agents. Through such communication, agents can obtain additional information, such as the internal logic of other controlled agents. This enables more accurate inference of other controlled agents' intentions and enhances their collaborative capabilities. Second, no communication should occur between controlled and uncontrolled agents. This is because the policies of uncontrolled agents are unknown, and thus their communication capabilities cannot be assumed in advance. Moreover, even if uncontrolled agents possess internal communication mechanisms, the ad hoc nature of the multi-agent system formation makes it impossible for the controlled agent system to fully comprehend their internal communication logic.

\begin{figure}[htbp]
\centering
\includegraphics[width=0.3\textwidth]{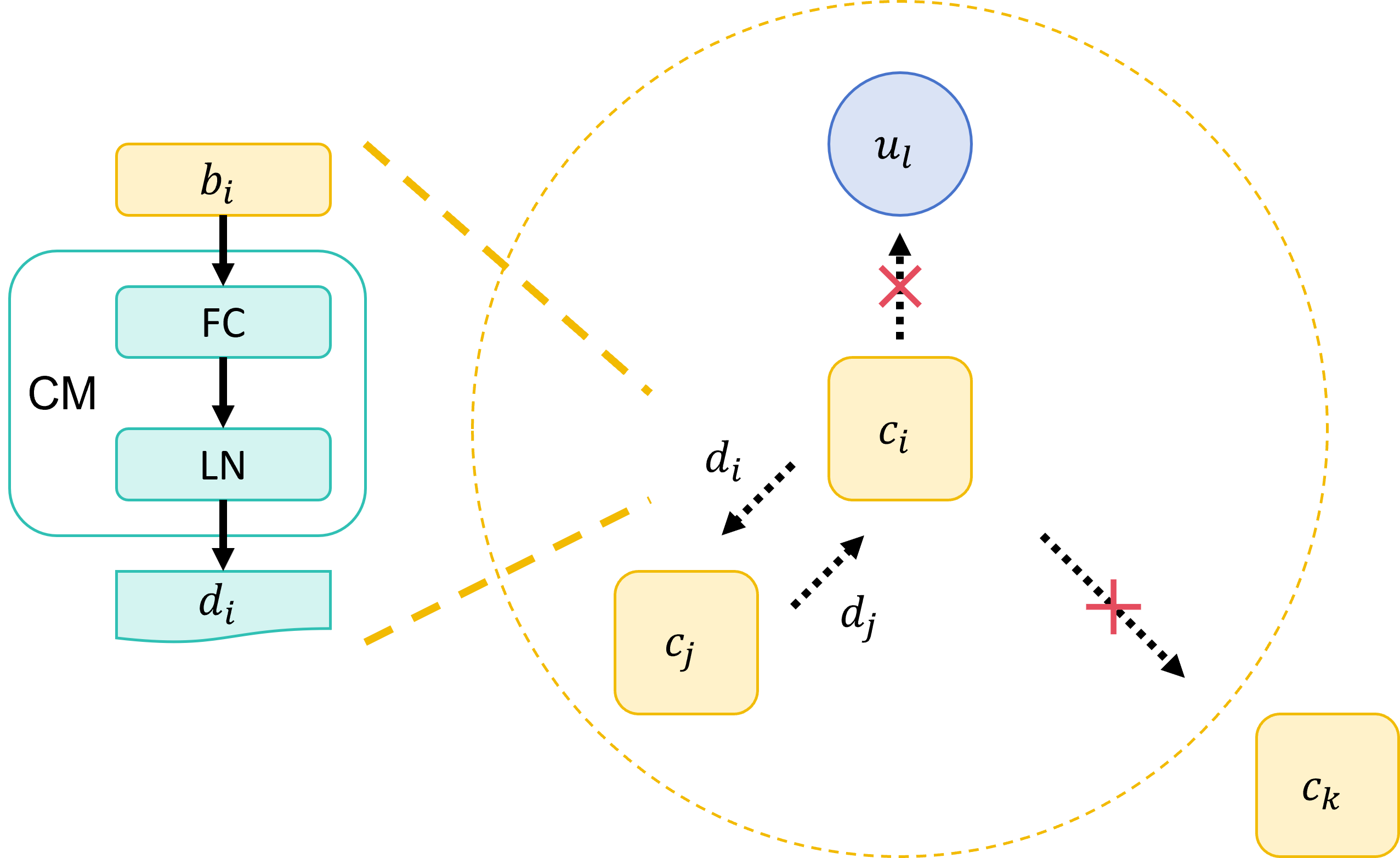}
\caption{Communication module (CM). The yellow dashed circle represents the communication range of the controlled agent $c_i$. According to CPCA, $c_i$ can only communicate with other controlled agent $c_j$ within its communication range. It cannot communicate with uncontrolled agent $u_l$, nor with any agent $c_k$ located outside its communication range. The communication data $d_i$ of a controlled agent $c_i$ is derived from its internal state $b_i$, processed through a fully connected layer (FC) and layer normalization (LN).}
\label{figure:communication}
\end{figure}

Figure~\ref{figure:communication} illustrates the principle of the communication module (CM). Since the internal state $b_i$ of a controlled agent $c_i$ is used to generate its policy $\pi_i$, it directly encapsulates the agent's intentions. In essence, the internal state $b_i$ represents the intrinsic mindset of agent $c_i$. Through the CM, the agent conveys its intentions via the communication data $d_i$, indirectly informing nearby controlled agents about its subsequent behavioral goals. This enables controlled agents to better understand each other's intentions, thereby strengthening their coordination and fostering mutual collaboration.

The generation of communication data can be expressed as
\begin{equation}
d_i = \text{LN}(\text{FC}(b_i)),
\end{equation}
where FC is a fully connected layer and LN is layer normalization. It is worth noting that the communication process itself has no dedicated loss function; its optimization occurs concurrently with the overall system optimization. This implies that the specific communication logic adopted among controlled agents cannot be directly inferred from external observations.

Indeed, an agent's behavior inherently conveys partial information about its internal intentions, allowing other agents to infer its goals through local observations. However, the local communication mechanism enables a more comprehensive expression of the agent's internal intentions, which will be experimentally validated.

\subsection{Information fusion}

The ability to process data of variable size is particularly crucial in open-system environments. When facing unknown settings and unfamiliar agents in such open systems, the number of local observations an agent can obtain is inherently unstable. For instance, there might be $k_t$ agents present in the local observations at the current timestep, while at the next timestep, this number may change to $k_{t+1}$. Therefore, learning to handle variable-length data can significantly enhance an agent's adaptability to open-system environments.

In terms of common existing approaches, a typical method involves determining the maximum data length based on the upper limit of the number of agents in the environment, and then padding the current data with zeros to a uniform length. However, the drawbacks of this method are evident. First, the padding process assigns a specific meaning to the padded values, which constrains the expressive capacity of the original data. Second, the order of the data can influence the final processing outcome. Third, once the number of data points exceeds the predefined upper limit, the method is forced to selectively discard some of the data.

\begin{figure}[htbp]
\centering
\includegraphics[width=0.35\textwidth]{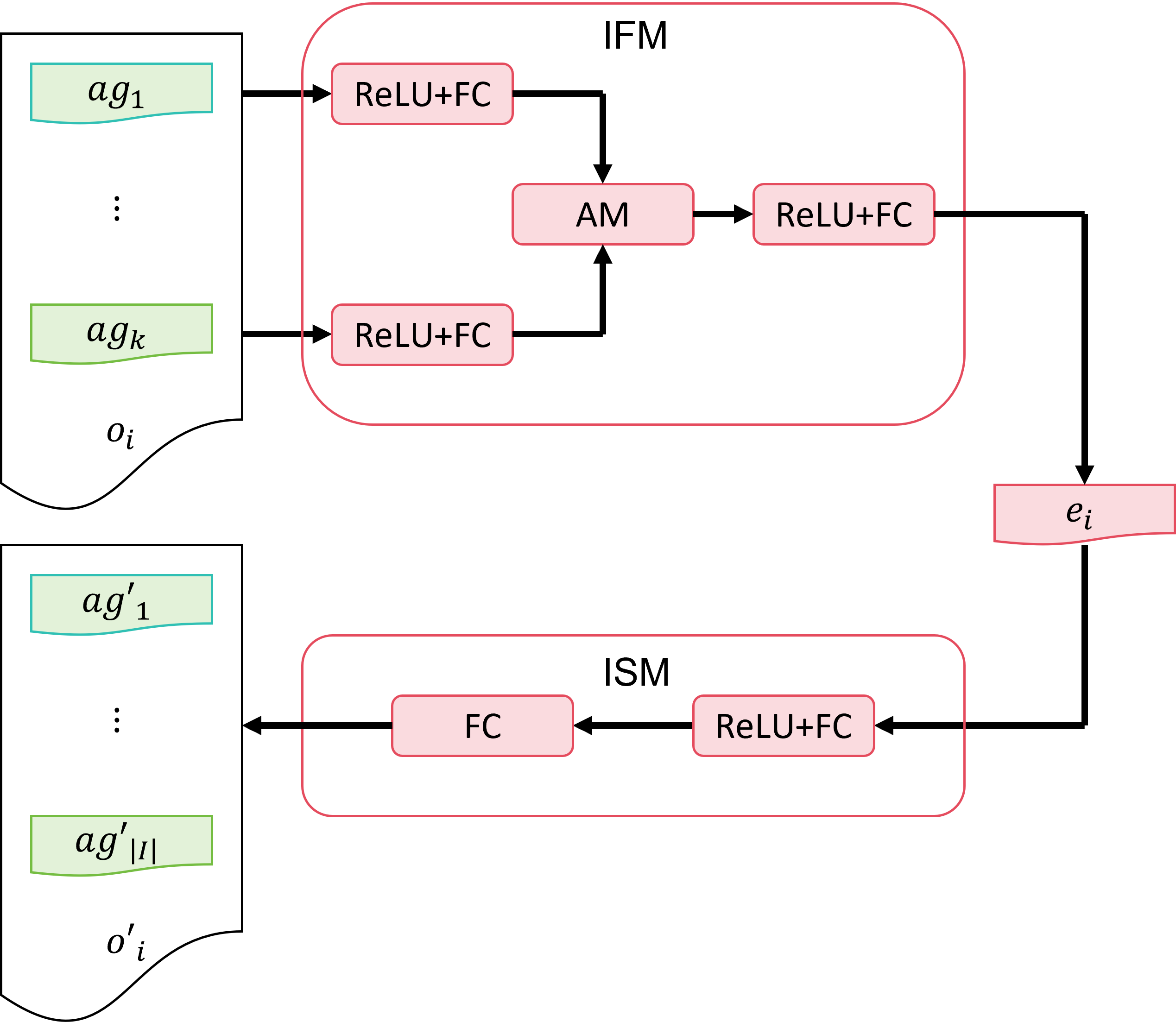}
\caption{Information fusion module (IFM) and information separation module (ISM). The controlled agent $c_i$ receives local observation information $o_i$, which contains data from $k$ agents $ag_1, \ldots, ag_k$. Each of these data entries is processed independently through an FC layer and a ReLU layer. An attention mechanism (AM) is then applied to fuse these processed data into a single vector. This vector is subsequently passed through another FC layer and ReLU activation to generate the representation $e_i$. During the training phase, $e_i$ is further processed through an FC layer, ReLU activation, and another FC layer to produce the predicted data $o'_i$. This predicted data $o'_i$ encompasses the predicted information of all agents $ag'_1, \ldots, ag'_{|I|}$.}
\label{figure:fusion}
\end{figure}

To partially address the aforementioned three limitations, we propose an information fusion module (IFM) and an information separation module (ISM), as illustrated in Figure~\ref{figure:fusion}. Within the IFM, the data preprocessing step employs parameter sharing; that is, the data $ag_j$ of each observed agent is processed by the same set of FC and ReLU layers. The data fusion is then accomplished through an attention mechanism (AM). This process can be represented as
\begin{equation}
\begin{split}
& \phi_j = \text{ReLU}(\text{FC}_1(ag_j)), \quad j=1,\ldots,k, \\
& [p_1,\ldots, p_k] = \text{softmax}[\eta^T \cdot \phi_1, \ldots, \eta^T \cdot \phi_k], \\
& e_i = \text{ReLU}(\text{FC}_2(p_1 \cdot \phi_1 + \ldots + p_k \cdot \phi_k + \eta )), \\
\end{split}
\end{equation}
where $\eta$ represents the trainable parameters of the AM.

This method partitions the observational data into atomic data, i.e., the data corresponding to each observed agent, and subsequently employs an AM to fuse these atomic data into a single vector. Notably, this approach avoids the use of padding to standardize data length, thereby preventing any numerical values from being assigned padding-specific meanings. This ensures that the representational capacity of the data remains uncompromised. Furthermore, since the AM integrates atomic data through summation, an operation that is commutative, the sequential order of the atomic data does not influence the final outcome. Additionally, the method does not presuppose an upper limit on the number of agents, allowing any finite set of atomic data to be fused into a unified vector via the AM.

During the training phase, the representation $e_i$ is also fed into the ISM to predict the data of all agents, i.e.
\begin{equation}
\begin{split}
& \psi = \text{ReLU}( \text{FC}_3 (e_i) ), \\
& [ag'_1,\ldots,ag'_{\vert I \vert}] = \text{FC}_4( \psi ). \\
\end{split}
\end{equation}
It is worth noting that since the maximum number of agents during training is fixed, the ISM does not need to handle variable-length data. In other words, padding can be used to standardize the data length. Furthermore, for the ISM, the order of agent data does not affect the training outcome. This is because the predicted data $ag'1, \ldots, ag'{|I|}$ is generated by $\psi$, and the structure of the FC layer is decomposable. That is, $[ag'_1, \ldots, ag'_{|I|}] = \text{FC}_4(\psi)$ is equivalent to
$$
ag'_j = \text{FC}_{4,j} (\psi), \quad j=1,\ldots, \vert I \vert.
$$
Therefore, changing the order of the predicted data $ag'_1, \ldots, ag'_{|I|}$ only affects the order of $\text{FC}_{4,1}, \ldots, \text{FC}_{4,|I|}$, with no impact on the preceding framework.

The ISM operates similarly to a spectral analysis of the representation $e_i$, decomposing it into predicted data for each agent. Therefore, during the training phase, the ISM requires a fixed upper limit on the number of agents and a predetermined order of predicted agents. By utilizing dedicated indicator bit to determine whether the data of a predicted agent is included in $e_i$, the meaning of padded values is governed by these indicators. Specifically, padded values are assigned their placeholder meaning only when the indicator bit signifies that the corresponding predicted agent's data is not contained in $e_i$.

For information fusion, it not only needs to be optimized along with the overall system but also has a separate reconstruction error term to optimize. For each controlled agent $c_i$, the reconstruction error term can be expressed as
\begin{equation}
L_{info, i} = \sum_{j=1}^{\vert I \vert} \Vert ag'_j - ag_{truth,j} \Vert^2,
\end{equation}
where $ag_{truth,j}$ represents the ground-truth data of the agent. This reconstruction error term indicates that the model training aims to make the predicted data approximate the ground-truth data more closely. Such an approach enhances the capability of the IFM to effectively fuse variable-length data while preventing information loss.

The proposed IFM/ISM belongs to an encoder-decoder architecture, yet differ significantly from conventional encoders and decoders. This distinction arises from our focus on enhancing agent adaptability in open-system environments, particularly through optimized information fusion of variable-length data. As a result, standard encoders are unsuitable for this scenario. Furthermore, integrating an AM into the encoder enables the multi-agent system to process acquired data more effectively, a capability that will be experimentally validated.

\section{Experiment} \label{section:experiment}

\subsection{Experimental setup}

StarCraft multi-agent challenge (SMAC) is a widely used benchmark environment for evaluating CMARL algorithms \cite{2019smac}. It is based on the popular real-time strategy game StarCraft II and focuses on micromanagement tasks where a team of decentralized agents must learn to control individual army units to defeat an opposing army controlled by the built-in game AI. 

\begin{table}[htbp]
\centering
\caption{Scenario description.}
\label{table:scenario}
\begin{tabular}{c c c c}
\toprule
 & Symmetry & & Asymmetry \\
\midrule
\multirow{2}{*}{Homogeneity} & \multirow{2}{*}{8m} & & 5m\_vs\_6m \\
 &  & & 8m\_vs\_9m \\
 & & & \\
\multirow{2}{*}{Heterogeneity} & 3s5z & & 3s5z\_vs\_3s6z \\
 & MMM & & MMM2 \\
\bottomrule
\end{tabular}
\end{table}

To comprehensively evaluate IFS through experiments, we selected a series of scenarios based on symmetry/asymmetry and homogeneity/heterogeneity criteria, as detailed in Table~\ref{table:scenario}. Symmetry refers to scenarios where opposing teams have identical agent configurations, while asymmetry typically describes situations involving a smaller team facing a larger one. A team is considered homogeneous if its agents share the same type and structure, and heterogeneous otherwise. It is worth noting that while the agents in 3s5z and 3s5z\_vs\_3s6z are heterogeneous, they all belong to the attack type. In contrast, the agents in MMM and MMM2 exhibit not only heterogeneity but also functional diversity, including both attack and healing types. Thus, although all four scenarios fall under the heterogeneous category, they emphasize distinct aspects of agent specialization.

\subsection{Performance comparison in NAHT problems}

\begin{table*}[htbp]
\centering
\caption{The average test return of different algorithms in the multi-agent ad hoc system.}
\label{table:performance}
\begin{tabular}{c c c c c c c c}
\toprule
Map & IPPO & IPPO-NAHT & QMIX & QMIX-NAHT & LIAM & POAM & IFS \\
\midrule
8m & 17.9 & 18.3 & 19.1 & 19.8 & 18.5 & 19.7 & \textbf{20.0} \\
5m\_vs\_6m & 10.8  & 14.3 & 11.6 & 13.4 & 11.0 & \textbf{15.2} & 15.1 \\
8m\_vs\_9m & 16.2 & 17.8 & 17.0 & 17.5 & 14.4 & 18.1 & \textbf{19.5} \\
3s5z & 14.1 & 15.6 & 14.8 & 18.4 & 18.1 & 19.9 & \textbf{20.1} \\
3s5z\_vs\_3s6z & 10.5 & 12.2 & 13.6 & 14.2 & 11.8 & 14.3 & \textbf{15.7} \\
MMM & 18.3 & 19.0 & 18.4 & 19.6 & 17.5 & 20.1 & \textbf{22.3} \\
MMM2 & 16.7 & 17.2 & 16.8 & 17.6 & 16.0 & 17.9 & \textbf{19.8} \\
\bottomrule
\end{tabular}
\end{table*}

We configure the uncontrolled agents using any one of the following algorithms: VDN \cite{2017value}, QMIX \cite{2018qmix}, IQL \cite{1997iql}, or IPPO \cite{2022ippo}. These uncontrolled agents do not participate in parameter optimization during experiments but instead serve as unknown agents in the NAHT setting. To validate our proposed IFS, we compare it against IPPO, IPPO-NAHT, QMIX, QMIX-NAHT, LIAM \cite{2021model}, and POAM \cite{2024adhoc} in SMAC experimental scenarios, as summarized in Table~\ref{table:performance}. It should be noted that among the baseline algorithms, both IPPO and QMIX treat NAHT as $n$ independent AHT problems. That is, these algorithms train only a single agent to cooperate with others, effectively reducing the multi-agent NAHT problem to $n$ separate single-agent AHT instances. When all $n$ agents are trained jointly to cooperate with others, the corresponding variants are IPPO-NAHT and QMIX-NAHT. LIAM, introduced in \cite{2021model}, essentially represents a degenerate case of POAM adapted for standard AHT settings.

The average test returns indicate that training a multi-agent ad hoc system as an integrated team generally yields superior performance compared to treating it as $n$ independent single-agent learners. This advantage arises because multi-agent systems inherently possess greater collective capacity than individual agents, enabling them to accomplish more complex tasks when coordinating with uncontrolled teammates. Furthermore, IFS achieves the highest average test returns in most scenarios, demonstrating the effectiveness of enhanced information flow and processing in multi-agent ad hoc systems. These results also reveal that existing multi-agent training methods like POAM do not fully realize the potential of agent coordination. By strengthening inter-agent relationships and facilitating communication, overall collaborative performance can be substantially improved. Moreover, in complex scenarios such as MMM2, the advantages of IFS through its deliberate information flow design become even more pronounced. In summary, establishing well-designed information flow and processing mechanisms based on informational principles can significantly enhance a system's capability to handle complex tasks.

\begin{figure*}[htbp]
\centering
\includegraphics[width=\textwidth]{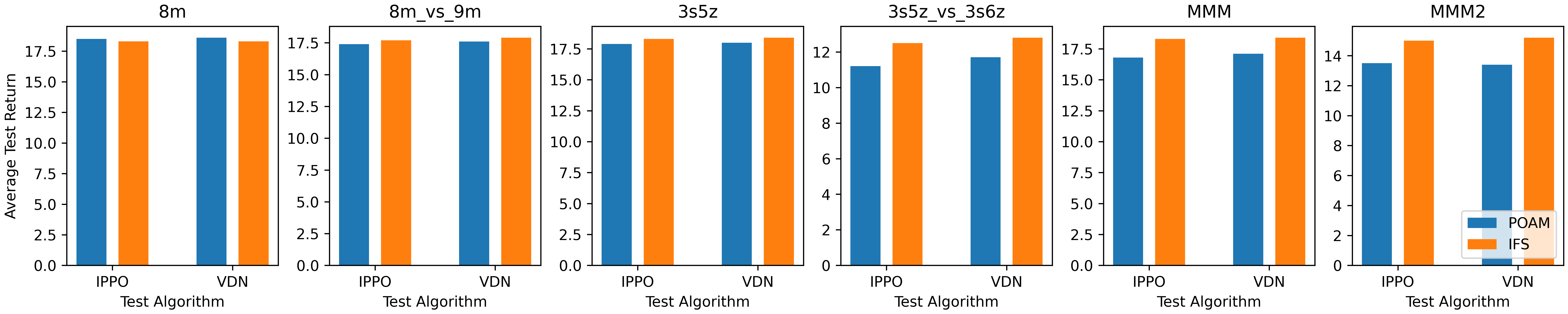}
\caption{Out-of-distribution (OOD) generalization testing. During training, the uncontrolled agent algorithms are selected from QMIX and IQL, while during testing, they are chosen from IPPO and VDN. Under general conditions, IFS demonstrates superior generalization performance compared to POAM, with this advantage being particularly pronounced in complex scenarios.}
\label{figure:ood}
\end{figure*}

To further evaluate the performance of IFS in NAHT problems, we employ out-of-distribution (OOD) generalization testing, where different algorithms are used for uncontrolled agents during training and testing phases. When the same algorithms are used for uncontrolled agents in both training and testing, this is referred to as in-distribution evaluation. However, to better align with real-world application scenarios, it is essential to test the generalization capability of IFS by employing previously unseen algorithms for uncontrolled agents during testing. For this purpose, we use QMIX and IQL as the algorithms for uncontrolled agents during training, and IPPO and VDN during testing, as illustrated in Figure~\ref{figure:ood}.

In the OOD generalization tests, both IFS and POAM exhibit some performance degradation. However, IFS generally demonstrates superior generalization capability compared to POAM, particularly in complex heterogeneous and asymmetric scenarios where it achieves higher average test returns. These results indicate that the deliberate design of information flow enhances the robustness of multi-agent ad hoc systems, enabling them to collaborate more effectively with unknown agents at the team level to accomplish tasks.

\subsection{Communication study}

\begin{figure*}[htbp]
\centering
\includegraphics[width=\textwidth]{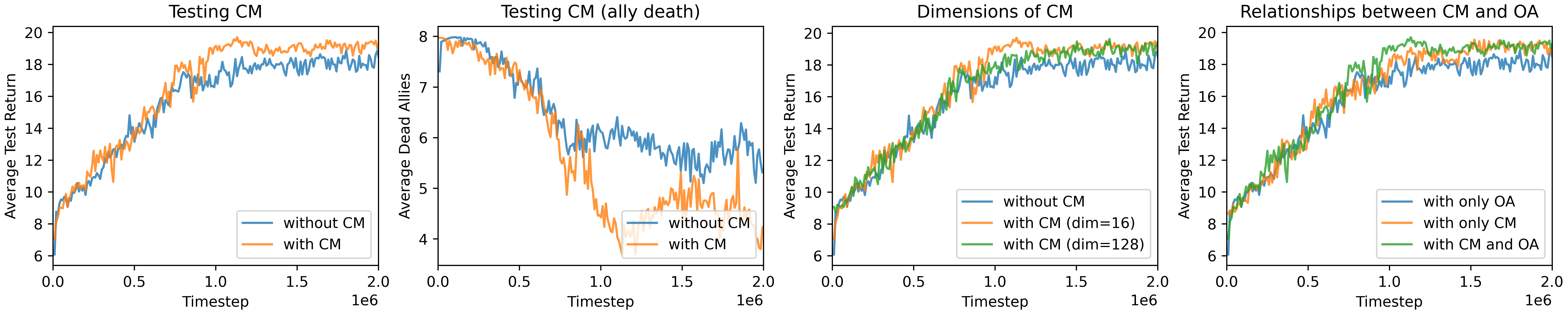}
\caption{Communication module testing. In the 8m\_vs\_9m scenario, we tested the performance of IFS with and without the CM. The testing CM experiment compares the average test returns and average ally deaths between the two configurations. Results show that the CM significantly enhances the ad hoc collaboration capability of controlled agents while reducing ally casualties. The dimensions of CM experiment examines the impact of varying communication dimensions, revealing that IFS exhibits low sensitivity to changes in this parameter. In the relationships between CM and OA experiment, we compared three information acquisition methods: communication-only (CM), observation of allies only (OA), and combined CM+OA. The findings demonstrate that CM alone is more effective than OA alone in improving cooperative performance.}
\label{figure:cm_result}
\end{figure*}

To evaluate the effectiveness of the CM designed based on CPCA, we conducted ablation studies on CM, experiments on communication dimensions within CM, and an investigation into the relationship between CM and observed allies (OA), as illustrated in Figure~\ref{figure:cm_result}. 

In the testing CM experiments, we conducted ablation studies by comparing configurations with and without the CM to evaluate its effectiveness. Results clearly demonstrate that incorporating the CM significantly enhances the performance of the multi-agent ad hoc system. Specifically, the CM effectively reduces ally casualties while achieving higher average test returns. These findings suggest that the CM strengthens inter-agent connections among controlled agents, enables more accurate inference of mutual intentions, and consequently improves collaborative capabilities when coordinating with uncontrolled teammates to complete tasks.

In the dimensions of CM experiments, we configured varying communication dimensions to investigate their impact on overall system performance. The results clearly indicate that the communication dimension in CM has limited influence on the performance of the multi-agent ad hoc collaboration system. While information theory dictates that the communication dimension cannot be excessively short, there exists a lower bound, excessively long dimensions do not enhance system performance and only increase training difficulty and computational burden. In summary, the multi-agent ad hoc collaboration system generally exhibits low sensitivity to variations in communication dimensions.

In the relationships between CM and OA experiments, we categorized the information acquisition methods among controlled agents into three types: OA only, obtaining observational data solely by observing other controlled agents; CM only, acquiring communication data exclusively through interaction with other controlled agents; CM+OA, utilizing both information sources. These three configurations were tested to investigate the relationship between CM and OA. Experimental results clearly show that the CM-only configuration achieves performance comparable to the CM+OA approach, while the OA-only configuration leads to a performance decline. This outcome indicates that information obtained through CM is richer than that acquired via OA. While OA typically provides conventional data such as position, type, and actions, CM conveys the intentions of other controlled agents, though it may also include some conventional information. Therefore, from an information flow perspective, CM effectively strengthens connections among controlled agents, thereby enhancing ad hoc collaborative capabilities.

\subsection{Information fusion study}

\begin{figure*}[htbp]
\centering
\includegraphics[width=\textwidth]{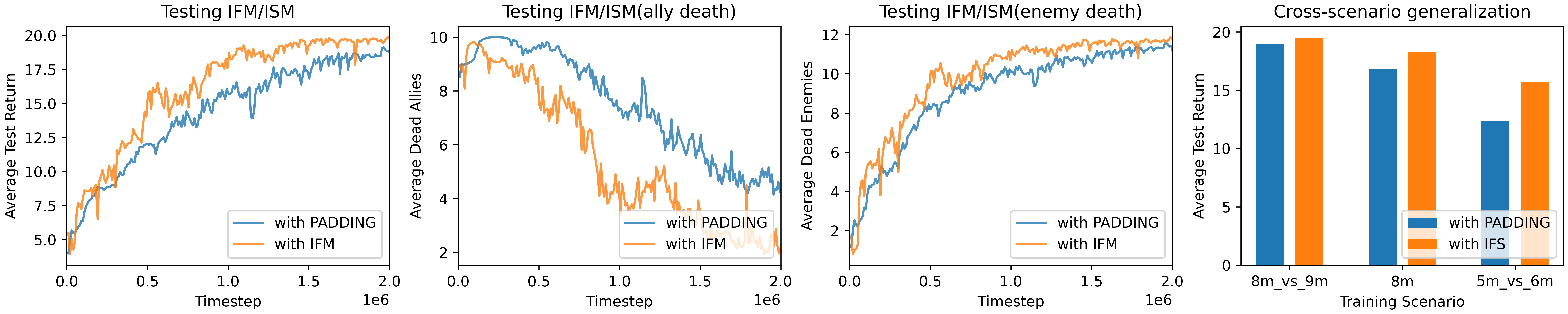}
\caption{Information fusion module / information separation module (IFM/ISM) testing. In the testing IFM/ISM experiments conducted in the MMM2 scenario, we compared two information processing approaches, using the IFM/ISM versus using padding (PADDING), to evaluate their impact on the multi-agent ad hoc system. The results demonstrate that IFM/ISM achieves higher average test returns, primarily by reducing ally casualties and increasing enemy eliminations. In the cross-scenario generalization experiments, agents trained in different environments (8m\_vs\_9m, 8m, 5m\_vs\_6m) were evaluated in the 8m\_vs\_9m test scenario to assess the generalization capability of IFM/ISM. The findings clearly indicate that IFM/ISM exhibits superior cross-scenario generalization performance.}
\label{figure:ifm_result}
\end{figure*}

To validate the effectiveness of the IFM/ISM, we conducted ablation studies on IFM/ISM and cross-scenario generalization experiments, as illustrated in Figure~\ref{figure:ifm_result}.

In the testing IFM/ISM experiments, we conducted a comparative analysis between systems utilizing the IFM/ISM and those employing the padding (PADDING) method. The results clearly demonstrate that IFM/ISM significantly enhances the average test return. This improvement is primarily attributed to IFM/ISM's ability to reduce ally casualties and increase enemy eliminations in the multi-agent ad hoc system. These findings confirm that IFM/ISM enables more effective information processing, consequently leading to more rational and coordinated strategies across the entire system.

In the cross-scenario generalization experiments, we decoupled the testing and training scenarios by evaluating multi-agent systems trained in different environments on a unified test scenario to validate the generalization capability of IFM/ISM. For the PADDING approach, when the number of data points obtained by controlled agents exceeds its processing limit, it discards part of the data. To create more meaningful tests, we trained multi-agent ad hoc systems in smaller scenarios and evaluated them in larger ones. Specifically, we trained systems in 8m\_vs\_9m, 8m, and 5m\_vs\_6m, then tested them all in the 8m\_vs\_9m scenario. Naturally, PADDING-based systems trained on 8m and 5m\_vs\_6m face data discarding issues when deployed in the larger test scenario. The experimental results show that while all methods experience performance degradation in unfamiliar scenarios, IFM/ISM consistently outperforms PADDING. This advantage primarily stems from IFM/ISM's ability to avoid data discarding by effectively integrating variable amounts of data. In contrast, PADDING inevitably loses information when data volume exceeds its predefined limit. Consequently, from an information flow perspective, IFM/ISM's superior handling of variable-length data directly translates to stronger cross-scenario generalization capability.

\subsection{Strategy visualization}

\begin{figure*}[htbp]
\centering
\begin{subfigure}[b]{0.24\textwidth}
\centering
\includegraphics[width=\textwidth]{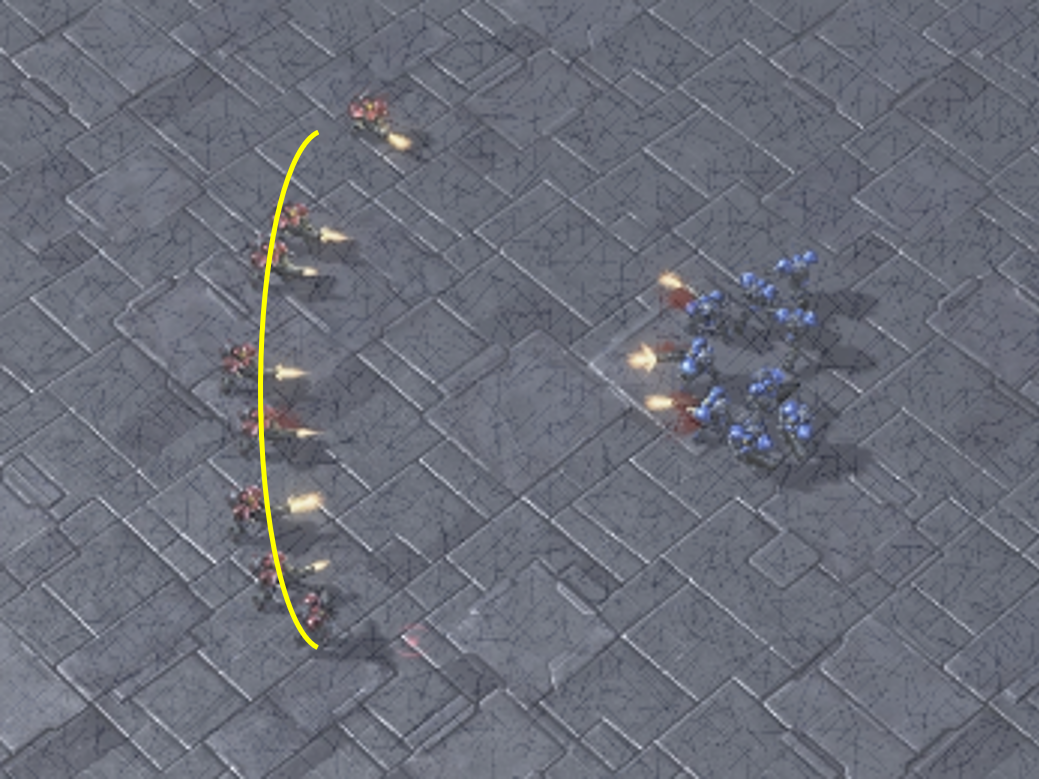}
\caption{}
\end{subfigure}
\hfill
\begin{subfigure}[b]{0.24\textwidth}
\centering
\includegraphics[width=\textwidth]{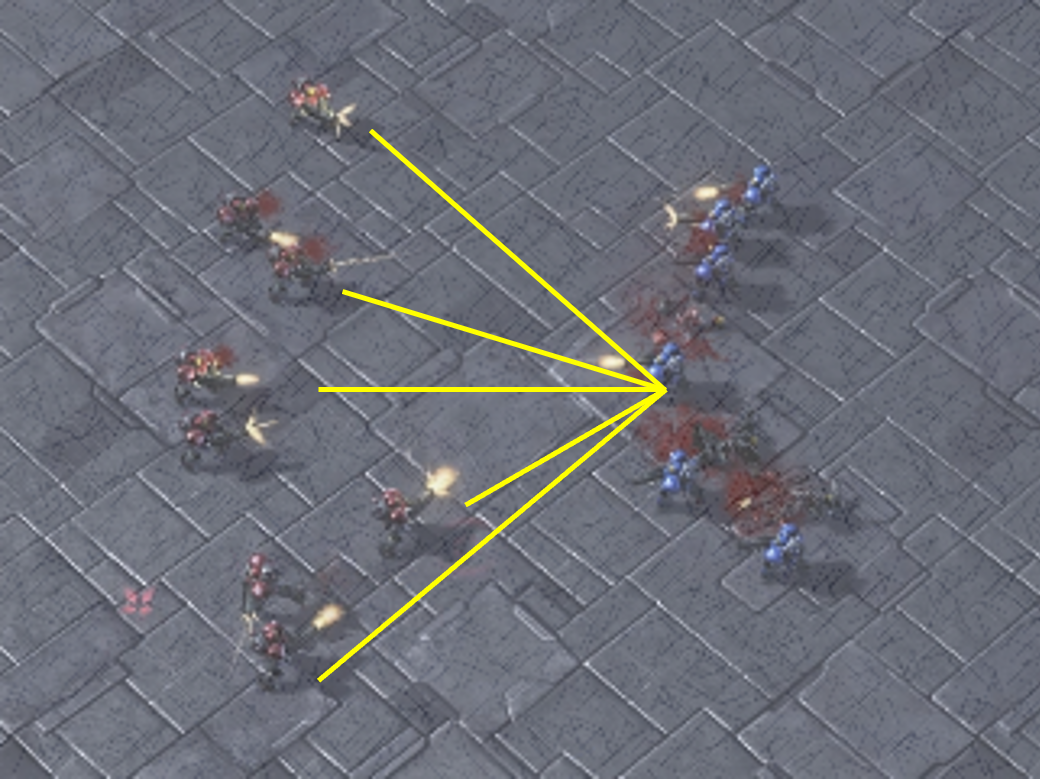}
\caption{}
\end{subfigure}
\hfill
\begin{subfigure}[b]{0.24\textwidth}
\centering
\includegraphics[width=\textwidth]{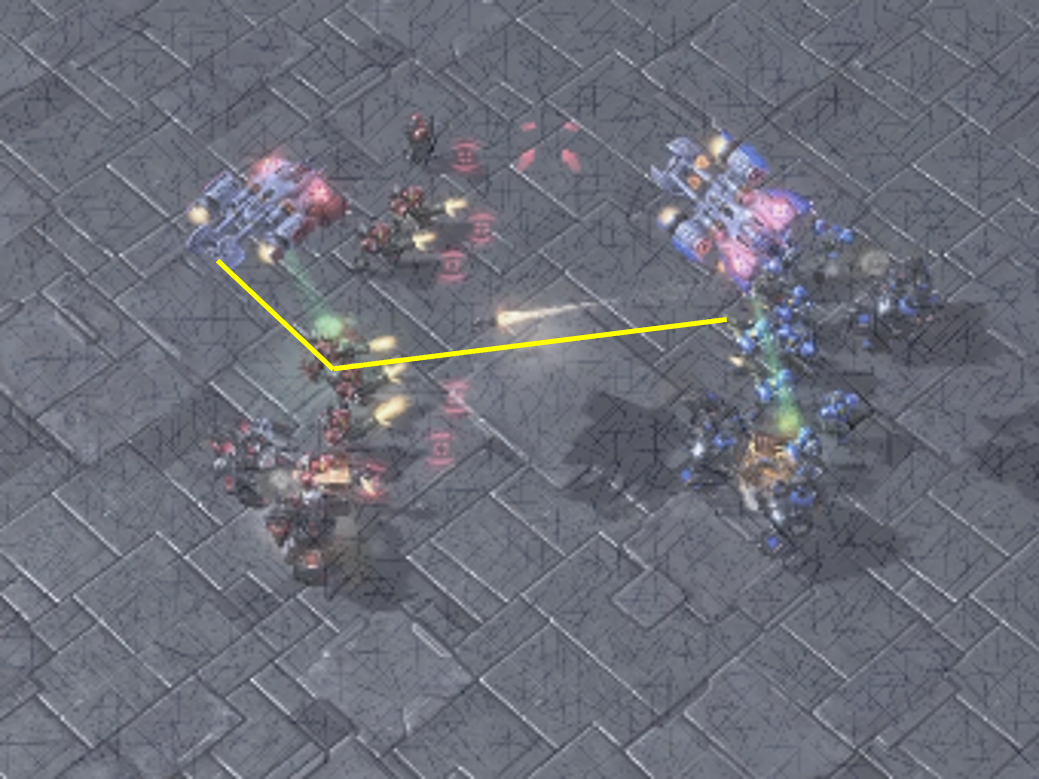}
\caption{}
\end{subfigure}
\hfill
\begin{subfigure}[b]{0.24\textwidth}
\centering
\includegraphics[width=\textwidth]{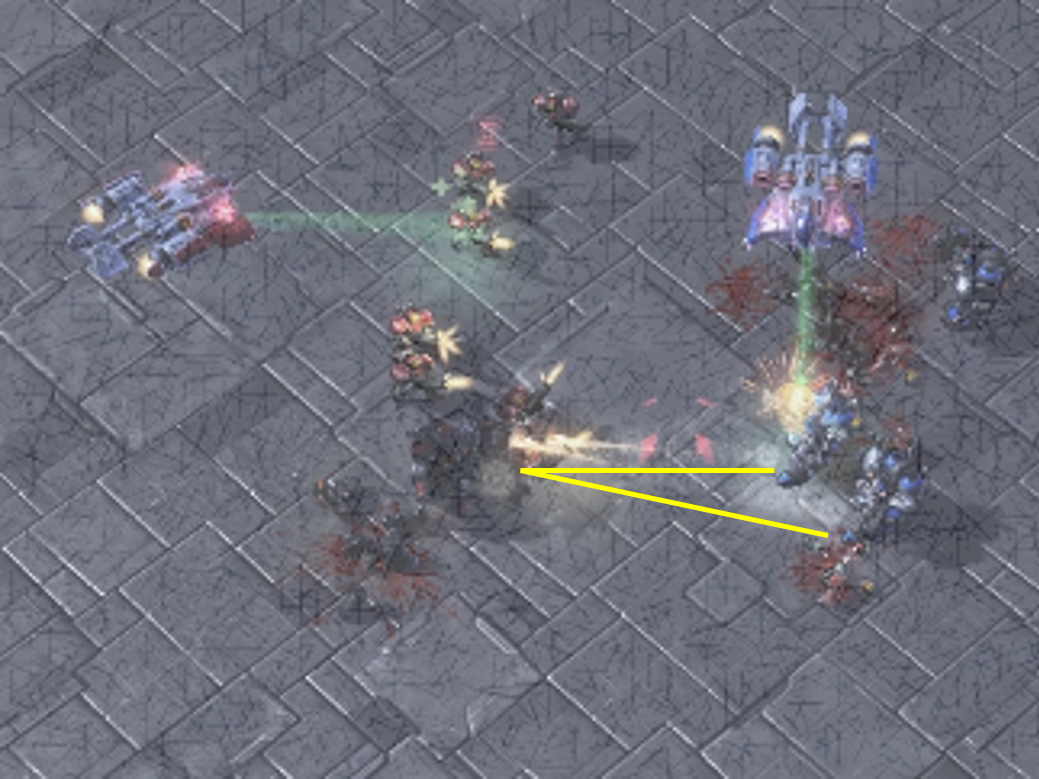}
\caption{}
\end{subfigure}

\caption{Strategy visualization. (a) Encirclement. (b) Concentrate fire. (c) Targeted healing. (d) Absorb damage.}
\label{figure:strategy}
\end{figure*}

To better demonstrate the strategies learned by the IFS-trained multi-agent ad hoc system, we visualize the multi-agent behaviors observed during experiments in Figure~\ref{figure:strategy}. Visually, these strategies align with established military doctrine, enabling the multi-agent ad hoc system to effectively coordinate with uncontrolled teammates in executing coordinated offensive operations.

\section{Conclusion} \label{section:conclusion}

Based on information flow analysis of multi-agent ad hoc systems, this paper identifies two key limitations in existing research: insufficient information flow and constrained information processing capacity. To address these issues, we propose an information flow structure for multi-agent ad hoc systems (IFS), tackling the two shortcomings through communication and information fusion perspectives respectively.

First, for the NAHT setting, we introduce a communication protocol for controlled agents (CPCA) and correspondingly design a communication module (CM) to strengthen internal coordination within the multi-agent ad hoc system and enhance collaboration capabilities with uncontrolled agents. Second, from the perspective of variable-length data processing, we propose an information fusion methodology and implement it through an information fusion module (IFM) and an information separation module (ISM). This approach eliminates the need for preset agent capacity limits, improves cross-scenario generalization performance, and makes controlled agents better suited for open-system environments. Finally, we systematically validate all proposed components through experimental scenarios in StarCraft II.

This paper primarily examines multi-agent ad hoc systems from an information flow perspective, and therefore employs identical network structures for all types of controlled agents to maintain simplicity. While specialized network architectures tailored to different agent types might potentially enhance the system's collaborative performance with uncontrolled teammates, representing a valuable and practically significant research direction, we leave this promising avenue for future investigation.

\bibliographystyle{ieeetr}
\bibliography{reference}

\begin{thebibliography}{10}

\bibitem{2018ai}
N.~Brown and T.~Sandholm, ``Superhuman ai for heads-up no-limit poker: Libratus
  beats top professionals,'' {\em Science}, no.~6374, p.~418–424, 2018.

\bibitem{2023tizero}
F.~Lin, S.~Huang, T.~Pearce, W.~Chen, and W.-W. Tu, ``Tizero: Mastering
  multi-agent football with curriculum learning and self-play,'' {\em
  Proceedings of the 2023 International Conference on Autonomous Agents and
  Multiagent Systems}, 2023.

\bibitem{2018qmix}
T.~Rashid, M.~Samvelyan, C.~Schroeder, G.~Farquhar, J.~Foerster, and
  S.~Whiteson, ``Qmix: Monotonic value function factorisation for deep
  multi-agent reinforcement learning,'' {\em Proceedings of the 35th
  International Conference on Machine Learning}, 2018.

\bibitem{1994markov}
M.~L. Littman, ``Markov games as a framework for multi-agent reinforcement
  learning,'' {\em Machine learning proceedings}, 1994.

\bibitem{2005cooperative}
L.~Panait and S.~Luke, ``Cooperative multi-agent learning: The state of the
  art,'' {\em Autonomous Agents and Multi-Agent Systems}, p.~387–434, 2005.

\bibitem{2019emergent}
B.~Baker, I.~Kanitscheider, T.~Markov, Y.~Wu, G.~Powell, B.~McGrew, and
  I.~Mordatch, ``Emergent tool use from multi-agent autocurricula,'' {\em
  International Conference on Learning Representations}, 2019.

\bibitem{2019graph}
J.~Jiang, C.~Dun, T.~Huang, and Z.~Lu, ``Graph convolutional reinforcement
  learning,'' {\em International Conference on Learning Representations}, 2019.

\bibitem{2019qtran}
K.~Son, D.~Kim, W.~J. Kang, D.~E. Hostallero, and Y.~Yi, ``Qtran: Learning to
  factorize with transformation for cooperative multi-agent reinforcement
  learning,'' {\em Proceedings of the 36th International Conference on Machine
  Learning}, p.~5887–5896, 2019.

\bibitem{2022aht}
R.~Mirsky, I.~Carlucho, A.~Rahman, E.~Fosong, W.~Macke, M.~Sridharan, P.~Stone,
  and S.~V. Albrecht, ``A survey of ad hoc teamwork research,'' {\em European
  Conference on Multi-Agent Systems}, p.~275–293, 2022.

\bibitem{2010aht}
P.~Stone, G.~Kaminka, S.~Kraus, and J.~Rosenschein, ``Ad hoc autonomous agent
  teams: Collaboration without pre-coordination,'' {\em Proceedings of the AAAI
  Conference on Artificial Intelligence}, vol.~24, p.~1504–1509, 2010.

\bibitem{2024adhoc}
C.~Wang, A.~Rahman, I.~Durugkar, E.~Liebman, and P.~Stone, ``N-agent ad hoc
  teamwork,'' {\em 38th Conference on Neural Information Processing Systems},
  2024.

\bibitem{2022decision}
A.~Kakarlapudi, G.~Anil, A.~Eck, P.~Doshi, and L.-K. Soh, ``Decision theoretic
  planning with communication in open multiagent systems,'' {\em Uncertainty in
  Artificial Intelligence}, p.~938–948, 2022.

\bibitem{2021coach}
B.~Liu, Q.~Liu, P.~Stone, A.~Garg, Y.~Zhu, and A.~Anandkumar, ``Coach player
  multi-agent reinforcement learning for dynamic team composition,'' {\em
  International Conference on Machine Learning}, p.~6860–6870, 2021.

\bibitem{2021open}
A.~Rahman, N.~Höpner, F.~Christianos, and S.~V. Albrecht, ``Towards open ad
  hoc teamwork using graph-based policy learning,'' {\em Proceedings of the
  38-th International Conference on Machine Learning}, vol.~139, 2021.

\bibitem{2023general}
A.~Rahman, I.~Carlucho, N.~HÃ¶pner, and S.~V. Albrecht, ``A general learning
  framework for open ad hoc teamwork using graph-based policy learning,'' {\em
  Journal of Machine Learning Research}, no.~298, p.~1–74, 2023.

\bibitem{2021formation}
Z.~Sui, Z.~Pu, J.~Yi, and S.~Wu, ``Formation control with collision avoidance
  through deep reinforcement learning using model-guided demonstration,'' {\em
  IEEE transactions on neural networks and learning systems}, p.~2358–2372,
  2021.

\bibitem{2019action}
W.~W, Y.~T, and L.~Y, ``Action semantics network: Considering the effects of
  actions in multiagent systems,'' 2019.

\bibitem{2021unmas}
C.~J, L.~W, and Z.~Y, ``Unmas: Multiagent reinforcement learning for unshaped
  cooperative scenarios,'' {\em IEEE transactions on neural networks and
  learning systems}, 2021.

\bibitem{2018modelfree}
Z.~Q, Z.~D, and L.~F. L, ``Model-free reinforcement learning for fully
  cooperative multi-agent graphical games,'' 2018.

\bibitem{2017value}
P.~Sunehag, G.~Lever, A.~Gruslys, W.~M. Czarnecki, V.~Zambaldi, M.~Jaderberg,
  M.~Lanctot, N.~Sonnerat, J.~Z. Leibo, and K.~Tuyls, ``Value-decomposition
  networks for cooperative multi-agent learning,'' 2017.

\bibitem{2021roma}
T.~Wang, H.~Dong, V.~Lesser, and C.~Zhang, ``Roma: Multi-agent reinforcement
  learning with emergent roles,'' {\em 37th International Conference on Machine
  Learning}, 2021.

\bibitem{2020multi}
D.~Mguni, J.~Wang, K.~Shao, L.~Chen, W.~Zhang, Y.~Yang, and Y.~Wen,
  ``Multi-agent determinantal q-learning,'' 2020.

\bibitem{2021qplex}
J.~Wang, Z.~Ren, T.~Liu, Y.~Yu, and C.~Zhang, ``Qplex: Duplex dueling
  multi-agent q-learning,'' {\em International Conference on Learning
  Representations}, 2021.

\bibitem{2021qvalue}
Y.~Yang, J.~Hao, G.~Chen, H.~Tang, Y.~Chen, Y.~Hu, C.~Fan, and Z.~Wei,
  ``Q-value path decomposition for deep multiagent reinforcement learning,''
  {\em 37th International Conference on Machine Learning}, 2021.

\bibitem{2021model}
G.~Papoudakis, F.~Christianos, and S.~V. Albrecht, ``Agent modelling under
  partial observability for deep reinforcement learning,'' {\em Advances in
  Neural Information Processing Systems}, 2021.

\bibitem{2021bayesian}
L.~Zintgraf, S.~Devlin, K.~Ciosek, S.~Whiteson, and K.~Hofmann, ``Deep
  interactive bayesian reinforcement learning via meta-learning,'' 2021.

\bibitem{2021trajectory}
A.~Lupu, B.~Cui, H.~Hu, and J.~Foerster, ``Trajectory diversity for zero-shot
  coordination,'' {\em International conference on machine learning}, 2021.

\bibitem{2018coma}
J.~Foerster, G.~Farquhar, T.~Afouras, N.~Nardelli, and S.~Whiteson,
  ``Counterfactual multi-agent policy gradients,'' {\em Proceedings of the AAAI
  conference on artificial intelligence}, vol.~32, 2018.

\bibitem{2021scaling}
F.~Christianos, G.~Papoudakis, M.~A. Rahman, and S.~V. Albrecht, ``Scaling
  multi-agent reinforcement learning with selective parameter sharing,'' {\em
  International Conference on Machine Learning}, p.~1989–1998, 2021.

\bibitem{2022ldsa}
M.~Yang, J.~Zhao, X.~Hu, W.~Zhou, J.~Zhu, and H.~Li, ``Ldsa: Learning dynamic
  subtask assignment in cooperative multi-agent reinforcement learning,'' {\em
  Advances in Neural Information Processing Systems}, vol.~35, p.~1698–1710,
  2022.

\bibitem{2020options}
A.~Vezhnevets, Y.~Wu, M.~Eckstein, R.~Leblond, and J.~Z. Leibo, ``Options as
  responses: Grounding behavioural hierarchies in multi-agent reinforcement
  learning,'' {\em International Conference on Machine Learning},
  p.~9733–9742, 2020.

\bibitem{2020other}
H.~Hu, A.~Lerer, A.~Peysakhovich, and J.~Foerster, ``“other-play” for
  zero-shot coordination,'' {\em International Conference on Machine Learning},
  p.~4399–4410, 2020.

\bibitem{2002Markov}
D.~S. Bernstein, R.~Givan, N.~Immerman, and S.~Zilberstein, ``The complexity of
  decentralized control of markov decision processes,'' {\em Mathematics of
  Operations Research}, no.~4, p.~819–840, 2002.

\bibitem{2023reinforcement-learning}
S.~E. Li, {\em Reinforcement Learning for Sequential Decision and Optimal
  Control}.
\newblock 152 Beach Road, 21-01/04 Gateway East, Singapore 189721, Singapore:
  Springer Nature Singapore Pte Ltd, 2023.

\bibitem{2019smac}
M.~Samvelyan, T.~Rashid, C.~S. De~Witt, G.~Farquhar, N.~Nardelli, T.~G.~J.
  Rudner, C.~M. Hung, P.~H.~S. Torr, J.~Foerster, and S.~Whiteson, ``The
  starcraft multi-agent challenge,'' 2019.

\bibitem{1997iql}
M.~Tan, ``Multi-agent reinforcement learning: Independent versus cooperative
  agents,'' {\em International Conference on Machine Learning}, 1997.

\bibitem{2022ippo}
C.~Yu, A.~Velu, E.~Vinitsky, Y.~Wang, A.~Bayen, and Y.~Wu, ``The surprising
  effectiveness of mappo in cooperative multi-agent games,'' {\em Proceedings
  of the Neural Information Processing Systems Track on Datasets and
  Benchmarks}, 2022.

\end{thebibliography}

\end{document}